\documentclass[conference]{IEEEtran}
\IEEEoverridecommandlockouts

\usepackage{todonotes}
\usepackage{cite}
\usepackage{amsmath,amssymb,amsfonts}
\usepackage{algorithmic}
\usepackage{graphicx}
\usepackage{textcomp}
\usepackage{xcolor}
\usepackage{subcaption}
\usepackage[acronym]{glossaries}

\def\BibTeX{{\rm B\kern-.05em{\sc i\kern-.025em b}\kern-.08em
    T\kern-.1667em\lower.7ex\hbox{E}\kern-.125emX}}

\newacronym{dm}{DM}{Decision Making}
\newacronym{mcc}{MCC}{Mobile Cloud Computing}
\newacronym{mec}{MEC}{Mobile Edge Computing}
\newacronym{cec}{CEC}{Cloud Edge Computing}
\newacronym{rsu}{RSU}{Road Side Unit}
\newacronym{uds}{UDS}{Unified Diagnostic Services}
\newacronym{sovd}{SOVD}{Service-Oriented Vehicle Diagnostics}
\newacronym{ecu}{ecu}{Electronic Control Unit}
\newacronym{dtc}{DTC}{diagnostic trouble codes}
\newacronym{obd}{OBD}{On-Board Diagnostics}
\newacronym{sae}{SAE}{Society of Automotive Engineers}
\newacronym{can}{CAN}{Controller-Area Network}
\newacronym{iso}{ISO}{International Organization for Standardization}
\newacronym{api}{API}{Application Programming Interface}

\usepackage[left=0.68in,right=0.673in,top=0.75in,bottom=1.05in]{geometry}

\begin{document}

\title{An OpenSource CI/CD Pipeline for Variant-Rich Software-Defined Vehicles\\
\thanks{Identify applicable funding agency here. If none, delete this.}
}

\author{
\IEEEauthorblockN{Matthias Weiß, Anish Navalgund, Johannes Stümpfle, Falk Dettinger and Michael Weyrich}
\IEEEauthorblockA{
\textit{Institute of Industrial Automation and Software (IAS)} \\
\textit{University of Stuttgart} \\
Pfaffenwaldring 47, 70550 Stuttgart, Germany \\
E-Mail: \{prename.surname\}@ias.uni-stuttgart.de}}

\maketitle

\begin{abstract}
Software-defined vehicles (SDVs) offer a wide range of connected functionalities, including enhanced driving behavior and fleet management. These features are continuously updated via over-the-air (OTA) mechanisms, resulting in a growing number of software versions and variants due to the diversity of vehicles, cloud/edge environments, and stakeholders involved. The lack of a unified integration environment further complicates development, as connected mobility solutions are often built in isolation. To ensure reliable operations across heterogeneous systems, a dynamic orchestration of functions that considers hardware and software variability is essential.
This paper presents an open-source CI/CD pipeline tailored for SDVs. It automates the build, test, and deployment phases using a combination of containerized open-source tools, creating a standardized, portable, and scalable ecosystem accessible to all stakeholders. Additionally, a custom OTA middleware distributes software updates and supports rollbacks across vehicles and backend services. Update variants are derived based on deployment target dependencies and hardware configurations.
The pipeline also supports continuous development and deployment of AI models for autonomous driving features. Its effectiveness is evaluated using an automated valet parking (AVP) scenario involving TurtleBots and a coordinating backend server. Two object detection variants are developed and deployed to match hardware-specific requirements. Results demonstrate seamless OTA updates, correct variant selection, and successful orchestration across all targets. Overall, the proposed pipeline provides a scalable and efficient solution for managing software variants and OTA updates in SDVs, contributing to the advancement of future mobility technologies.
\end{abstract}


\section{Introduction}\label{sec:introduction}

Software-defined vehicles (SDVs) represent a transformative shift in the automotive industry, enabling dynamic, software-driven functionalities that enhance vehicle behavior, safety, and user experience \cite{goswami2024sdv} \cite{dettinger2024function}. Unlike traditional vehicles with fixed capabilities, SDVs leverage continuous connectivity to cloud and edge infrastructures to enable advanced features such as real-time traffic-based routing, predictive maintenance, and remote fleet management \cite{madhuri2025sdv} \cite{dettinger2024survey}. This software-centric evolution necessitates frequent and often automated software updates via over-the-air (OTA) mechanisms, making vehicles not only transport devices but also continuously evolving cyber-physical systems \cite{weiss2023continuous}. 

A central challenge emerging from this evolution is the growing complexity in managing software variants. Even conventional vehicles exhibit significant variability due to regional regulations, hardware configurations, and feature bundles. With the increasing reliance on software, this variability expands dramatically: different vehicle generations, sensor configurations, compute platforms, and regional software stacks must be considered \cite{seidel2024variant}. Consequently, each update potentially requires multiple tailored variants to ensure compatibility and functionality across a heterogeneous fleet.

To address this complexity, continuous integration and continuous deployment (CI/CD) pipelines have become a critical tool in modern software engineering. These pipelines automate the build, testing, and deployment of software, supporting rapid and reliable iterations. In the context of SDVs, CI/CD practices promise to streamline the development and delivery of both onboard and backend services, ensuring consistent quality and faster rollout of innovation\cite{weiss2024simulating}.

However, despite the availability of mature CI/CD tools, two key challenges remain unresolved in the automotive domain. First, SDVs require \textit{stable orchestration of updates across distributed targets}—including in-vehicle platforms, edge nodes, and backend cloud services—each with unique requirements and availability constraints \cite{federate2024roadmap}. Second, the \textit{management of software variants}—where a single codebase must be adapted to different hardware configurations and software environments—requires intelligent deployment logic and variant-aware tooling \cite{stoetzner2022deployment}.

In light of these challenges, this paper addresses the following research question:

\textit{How can an open-source CI/CD pipeline be designed to support stable, variant-aware deployment of connected vehicle software across heterogeneous targets including cloud, edge, and in-vehicle environments?}

The remainder of this paper is structured as follows: Section~\ref{sec:relatedwork} reviews related work and existing CI/CD solutions in automotive and distributed systems. Section~\ref{sec:architecture} presents the architecture and components of the proposed open-source CI/CD pipeline. Section~\ref{sec:evaluation} evaluates the pipeline using an automated valet parking (AVP) scenario.  
Finally, Section~\ref{sec:conclusion} concludes the paper and outlines directions for future research.

\section{Background}\label{sec:relatedwork}

\subsection{Variant-rich Software-Defined Vehicles}
The high variability in software-defined vehicles enables unprecedented configurability and individuality, allowing each vehicle to be tailored to specific market requirements and customer preferences. However, this same flexibility introduces substantial complexity in managing the resulting software ecosystem \cite{wozniak2015automotive}. With SDVs designed to remain current through continuous over-the-air updates throughout their extended lifecycle—often spanning 10-15 years—the combination of high variance and frequent updates creates a significant risk of software erosion \cite{stumpfle_tackling_2024}.

This risk is particularly concerning given that SDVs fundamentally rely on software for their core functionality. To address this challenge, the automotive industry is increasingly adopting software product line engineering (SPLE) approaches \cite{knieke_managed_2022}. SPLE provides systematic methods for managing commonalities and variabilities across the vehicle software portfolio, enabling efficient development and maintenance of multiple product variants from a shared set of core assets. The challenge however lies in adopting such an SPL \cite{10710832,11024511}. As vehicles receive regular updates to add features, improve performance, and address security vulnerabilities, the number of possible software configurations grows exponentially. Each update must be compatible not only with the base vehicle platform but also with the unique combination of features and variants present in individual vehicles.
Modern middleware solutions like Adaptive AUTOSAR incorporate SPLE principles, providing service-oriented architecture for modular updates and Update and Configuration Management (UCM) services \cite{noauthor_autosar_nodate}. 
The challenge extends beyond technical deployment to encompass version control, dependency management, and regression testing across an ever-expanding matrix of software variants.

\subsection{DevOps and CI/CD}
DevOps is a software engineering paradigm that emphasizes collaboration between development and operations teams, supported by extensive automation. Through continuous integration (CI) and continuous deployment (CD), DevOps enables rapid, reliable software updates. In the automotive domain, this approach is increasingly relevant as software-defined vehicles demand frequent OTA updates to both onboard functions and connected backend services \cite{weiss2022devops}.

CI/CD pipelines implement DevOps principles by automating build, test, and deployment processes. With the growing role of machine learning (ML) in SDVs—e.g., for perception or driver behavior—MLOps has emerged as an extension of CI/CD to support continuous training, validation, and deployment of ML models \cite{ferreira2025mlops}. Together, CI/CD and MLOps provide a foundation for agile, data-driven vehicle software development.

Recent industry initiatives such as COVESA \cite{covesa2025web} and SOAFEE \cite{soafee2025doc} promote cloud-native architectures and runtime environments for vehicles, bringing containerization and orchestration tools to the edge. Academic efforts have demonstrated CI/CD integration with physical vehicles, enabling rapid iteration and feedback loops. However, two major gaps persist. First, variant management remains a challenge: software must be tailored to diverse vehicle configurations, yet current tools lack native support for variant-aware deployment \cite{stoetzner2022deployment}. Second, constructing closed ML data loops—where field data informs continuous model improvement and redeployment—is difficult due to bandwidth, validation, security and integration constraints \cite{federate2024roadmap} \cite{weiss2024ad}.

Overall, while the foundations for DevOps in SDVs exist, an end-to-end CI/CD solution that unifies backend orchestration, in-vehicle deployment, variant management, and ML lifecycle integration is still lacking.
  
\section{Pipeline Architecture}\label{sec:architecture}
\subsection{Overview}
\begin{figure*}[htp]
    \centering
    \includegraphics[width=\textwidth]{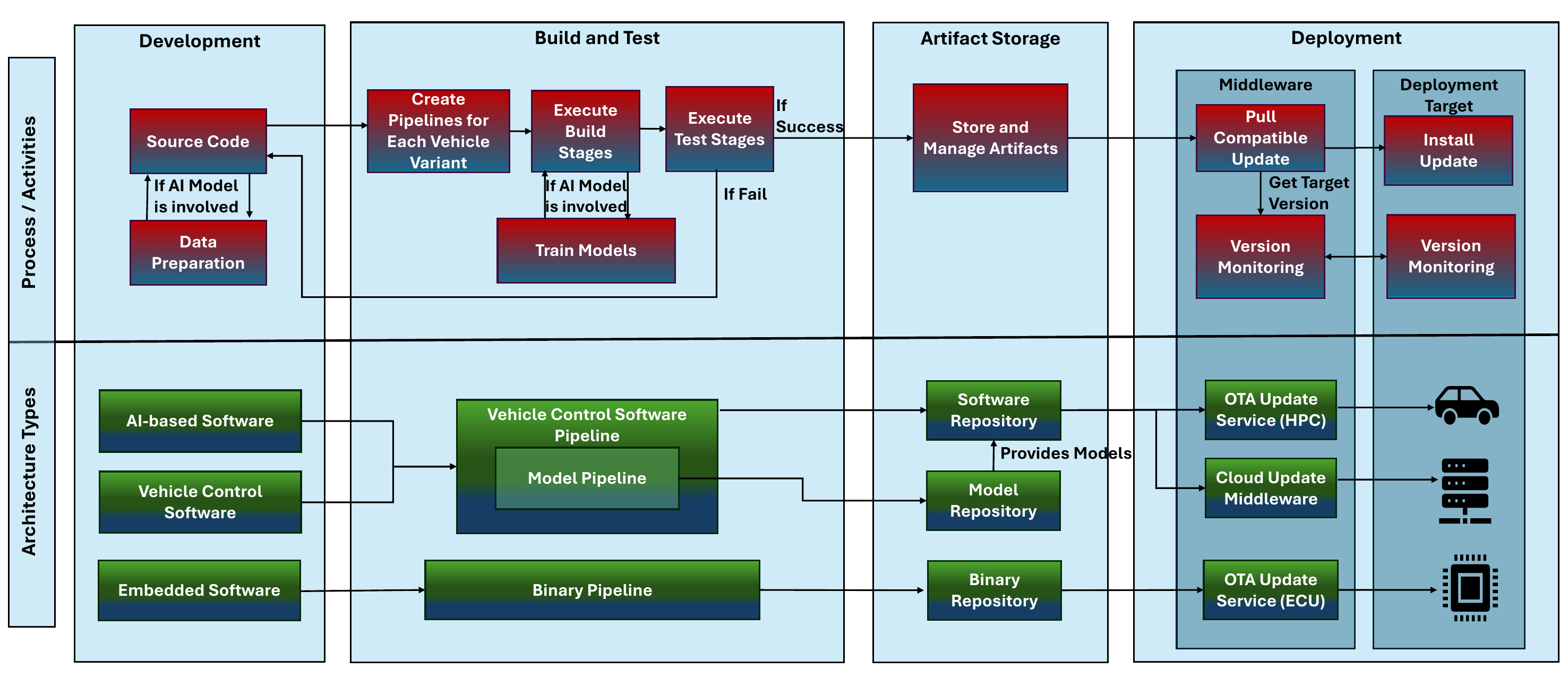}
    \caption{Architecture of the CI/CD pipeline, subdivided into phases and artifacts.}
    \label{fig:architecture}
\end{figure*}

To address the research question and gaps presented in Section \ref{sec:introduction} and \ref{sec:relatedwork}, this Section proposes the CI/CD pipeline architecture shown in Fig.~\ref{fig:architecture} as a modular and extensible framework for managing software development, integration, and deployment across variant-rich software-defined Vehicles. The system distinguishes between three different software types, namely embedded firmware, vehicle control software built as services, and AI-based components such as machine learning models. The architecture is structured into four core functional domains, following conventional CI/CD processes: Development, Build and Test, Artifact Storage, and Deployment.

In the Development phase, the process begins with source code and optionally includes data preparation when AI models are involved. Variant-specific pipelines are instantiated based on target configurations, enabling tailored workflows for each vehicle class or Electronic Control Unit (ECU).

The Build and Test stage executes dedicated pipelines for each software category. Build stages include compilation, containerization, or model training, followed by automated test stages. Failures are flagged for reprocessing, while successful builds are forwarded to storage.

The Artifact Storage layer manages validated software artifacts using type-specific repositories: binaries, containers, and trained models are stored and indexed for traceability and version control. These repositories serve as the source of truth for deployment.

Finally, the Deployment phase involves both cloud-side and client-side OTA middleware. The cloud component determines target compatibility, retrieves artifacts from storage, and coordinates update rollouts. The client middleware, deployed on the vehicle’s hardware, handles version monitoring, artifact validation, and update installation. The architecture supports hierarchical deployment across heterogeneous compute units, including embedded ECUs, High Performance Computers (HPCs), and backend services.

Full details on all process steps are provided in the following.

\subsection{Development Phase}



The development phase 
marks the entry point of the CI/CD pipeline and is responsible for managing the initial software source code and data preparation tasks. The architecture supports multiple software modalities—embedded software, vehicle control software, and AI-based models—each of which originates from a shared or modularized codebase. Based on the target vehicle variant, the pipeline dynamically creates specific build configurations using pre-defined templates or variant descriptors.

When AI software is involved, a data preparation block is activated before the pipeline proceeds. This step includes dataset retrieval, versioning, annotation, and pre-processing required for training or updating a machine learning model. The outcome of this step is a curated dataset ready for training or retraining within the build stage.

The key outcome of the development phase is the generation of variant-specific build pipelines, each of which encapsulates the necessary configuration to process one or more software types based on the deployment target. These pipelines ensure that dependencies, build logic, and testing parameters are tailored to the vehicle’s hardware profile, supported sensors, and compute capabilities.

By separating software types and integrating optional AI data flows, the development block creates a flexible foundation for supporting SDV heterogeneity without duplicating build logic. This abstraction also enables parallel development tracks and faster iteration cycles.

\subsection{Build and Test Phase}


The Build and Test phase 
processes the variant-specific pipelines generated during development. This phase is responsible for compiling source code, training machine learning models, and executing validation routines tailored to the target artifact type—embedded binaries, vehicle control logic, or AI-based models.

Each pipeline begins with the execution of build stages. For embedded software, this includes cross-compilation and linking for the target ECU architecture. For vehicle control software, container images are constructed using Docker or Snapcraft. If the software includes machine learning components, model training is executed using the preprocessed dataset and specified hyperparameters. The full details on the MLOps pipeline are described in Section \ref{sec:mlops}. All build jobs are monitored for success or failure.

Upon successful build completion, the pipeline proceeds to automated test stages. These include unit tests, integration tests, simulation runs, and model validation metrics, depending on the software category. If any stage fails, the build is rejected and flagged for correction. Successful outputs are forwarded to the artifact storage stage for deployment.

By decoupling build logic per software type and incorporating conditional model training, this phase ensures that all artifacts are rigorously validated before delivery, supporting software reliability and functional safety in SDVs.

\subsection{Storage Phase}
The Artifact Storage phase acts as a persistent and versioned repository layer for all validated software artifacts. After successful completion of build and test stages, artifacts are classified by type and pushed to their respective storage systems. Three distinct repositories are used:
\begin{itemize}
    \item The Binary Repository stores low-level firmware and embedded executables, often in formats suitable for direct flashing onto ECUs.
    \item The Model Repository handles trained AI models, including weights, configuration files, and metadata such as version, accuracy, and target hardware compatibility.
    \item The Software Repository stores containerized applications or middleware packages, typically in Docker or Snap formats, used for vehicle control or backend services.
\end{itemize}
Each repository supports artifact versioning, access control, and metadata tagging, enabling precise selection during deployment. For AI-based pipelines, the Model Repository can also feed into the Software Repository when inference components are embedded into deployable containers. This cross-referencing ensures the correct AI models are packaged with the control software.

By separating repositories by artifact type, the architecture supports flexible, version-aware deployment strategies while maintaining traceability and compliance for safety-critical applications.

\subsection{Deployment Phase}


The Deployment phase in the proposed architecture consists of two key components: OTA Middleware and Deployment Targets. Together, they enable secure, scalable, and variant-aware delivery of software artifacts across the fleet. 
 
\subsubsection{OTA Middleware}
The OTA middleware is responsible for coordinating the end-to-end update process. It is divided into two logical layers:

\textbf{Cloud Middleware:} Deployed on backend infrastructure, the cloud middleware continuously monitors versioned artifact repositories for changes. It determines artifact-to-vehicle compatibility based on metadata such as software type, variant configuration, and hardware specifications. It manages rollout strategies, including staged deployment, canary updates, and full-fleet upgrades. For AI-based software, it verifies that model weights are suitable for the vehicle's sensors and compute capabilities. Once validated, the appropriate artifact version is queued for deployment.

\textbf{Client Middleware:} This component resides on the vehicle itself, either on the HPC or a central ECU responsible for the update process. It periodically polls the cloud middleware for updates, verifies version compatibility, retrieves the relevant artifact, and installs it locally. Post-installation, it performs integrity checks and provides telemetry back to the cloud for monitoring and rollback support.

\subsubsection{Deployment Targets}
Once updates are processed through the OTA middleware, they are routed to specific runtime targets based on software type:

\textbf{ECUs} receive embedded binaries, often as compiled firmware, through flashing procedures managed by the client middleware.

\textbf{HPCs} run vehicle control software as containerized applications, typically deployed via Docker or Snap. If applicable, trained AI models are mounted within the containers.

The \textbf{cloud backend} executes connected vehicle functions and allows for function offloading of resource-heavy tasks into the cloud. 

Each target environment includes built-in support for version tracking, artifact validation, and rollback handling. This ensures that updates are not only correctly delivered but also verified and monitored during execution.

\subsection{MLOps Data Pipeline}\label{sec:mlops}

\begin{figure}[htp]
    \centering
    \includegraphics[width=\columnwidth]{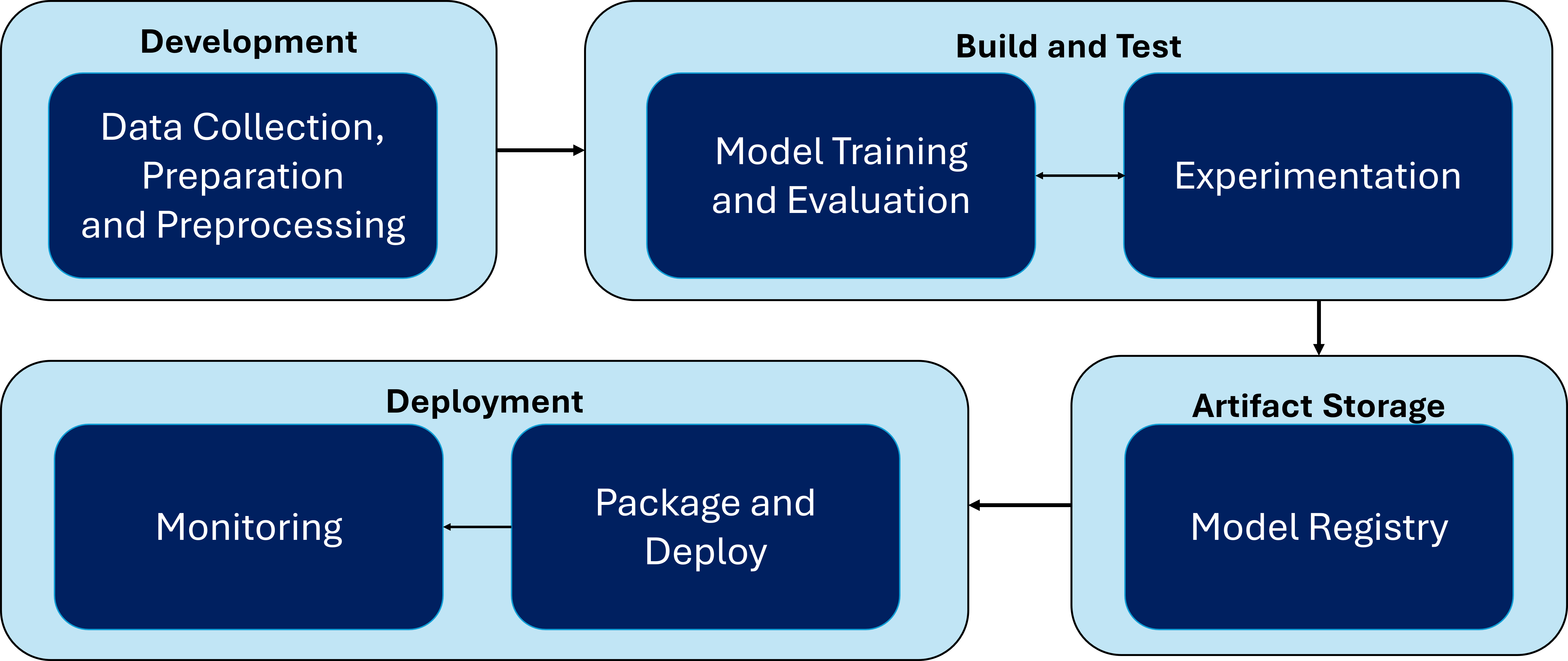}
    \caption{MLOps Data Pipeline for AI model deployment in SDVs}
    \label{fig:aidata}
\end{figure}

Fig.~\ref{fig:aidata} shows the MLOps pipeline which enables the structured development, integration, and deployment of machine learning models in variant-rich SDVs. 

The MLOps data loop in the proposed architecture is integrated into the CI/CD pipeline and, equivalent to the regular pipeline, organized into four logical phases: Development, Build and Test, Artifact Storage, and Deployment. This structure enables continuous development, evaluation, and deployment of machine learning models tailored to SDVs.

In the Development phase, data is collected from vehicle sensors such as cameras, LiDAR, or accelerometers. This data is versioned for traceability and subsequently processed in two steps: data preparation (including annotation and splitting) and data preprocessing (such as resizing, normalization, or augmentation). 

The Build and Test phase involves supervised model training using the prepared data. This step includes hyperparameter tuning and evaluation on validation datasets. A parallel experimentation block supports iterative testing of different model architectures. Once a model passes evaluation criteria, it is pushed to the Model Registry.

In the Artifact Storage phase, the Model Registry serves as a version-controlled repository for validated models. It stores associated metadata, performance metrics, and compatibility constraints required for downstream deployment.

The Deployment phase pushes the selected model to the appropriate runtime environment, such as an on-vehicle inference node. The deployed model is continuously monitored for accuracy, latency, and anomalies.

This pipeline ensures that machine learning components in SDVs remain reliable, up-to-date, and adaptive to changing real-world conditions, while integrating seamlessly with the overall CI/CD workflow.

\section{Evaluation} \label{sec:evaluation}
A laboratory testbed was constructed to evaluate the proposed CI/CD pipeline for AI software, vehicle control software, and embedded binary software deployment. The setup reflects a realistic SDV environment and includes multiple TurtleBots connected to HPCs and a centralized backend server hosting the CI/CD infrastructure, OTA middleware, and artifact repositories. In addition, ESP32-based devices were used to emulate Electronic Control Units (ECUs) for evaluating low-level binary software deployment. In the following, the evaluation setups and results for all software types are presented.

\subsection{Updating Embedded Software}
\begin{figure}[htp]
\centering
\includegraphics[width=\columnwidth]{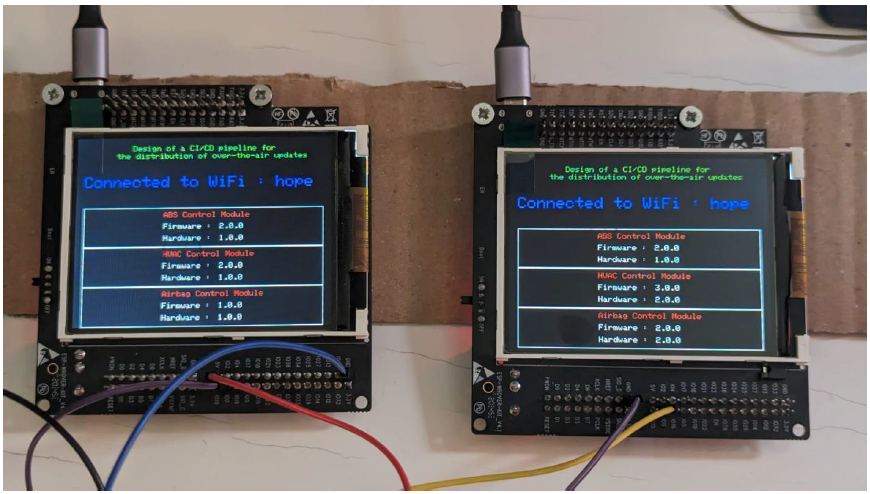}
\caption{Vehicle variant 1 (left), Vehicle variant 2 (right)}
\label{fig:binary_result}
\end{figure}

To assess embedded binary deployment, two vehicle variants with different ECU configurations were used. Variant-specific OTA packages were generated based on a vehicle dependency matrix defining firmware compatibility for each ECU. Over-the-Air (OTA) updates were executed on both vehicle variants using the OTA middleware. Each vehicle received a variant-specific OTA package, as determined by the vehicle dependency matrix. Fig.~\ref{fig:binary_result} illustrates the results of the update process. While the hardware versions of the ECUs remained unchanged, the firmware versions differed between the variants. Vehicle Variant 1 received ECU versions 2.0.0, 2.0.0, and 1.0.0 for the ABS Control Module, HVAC Control Module, and Airbag Control Module, respectively. Vehicle Variant 2 was updated with versions 2.0.0, 3.0.0, and 2.0.0 for the corresponding modules.

To validate rollback functionality, a software rollback was performed on both vehicle variants. Distinct rollback packages were pre-stored on the Artifactory server for each variant. Using the OTA middleware, the vehicles successfully retrieved and installed the correct rollback builds based on their variant configuration, restoring their ECUs to the prior firmware state.

\subsection{Updating Vehicle Control Software}
TurtleBot 4 robots were used for evaluating vehicle control and AI-based software. They serve as an open-source mobile robot platform running ROS 2 and include various onboard sensors, such as 2D LiDAR, an OAK-D stereo camera, infrared, and bump detection sensors, supporting applications like autonomous navigation and perception.

\begin{figure}[htp]
\centering
\includegraphics[width=\columnwidth]{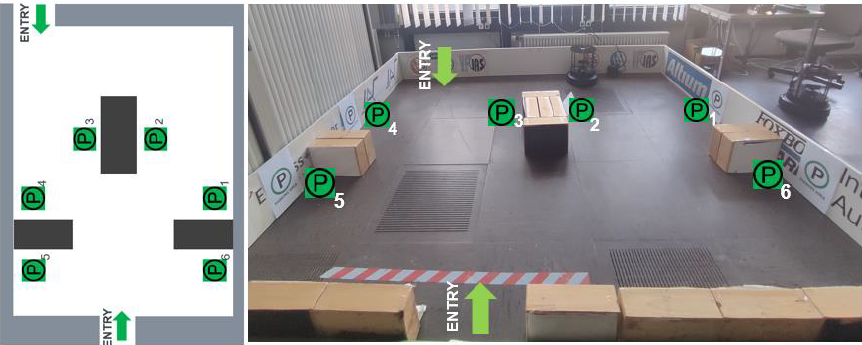}
\caption{Laboratory setup for Autonomous Valet Parking}
\label{fig:vp}
\end{figure}

Inside the lab setup, an autonomous valet parking (AVP) software was developed, enabling TurtleBots to navigate to the closest parking spot in a lab course. The complete setup can be seen in Fig.~\ref{fig:vp}. Multiple parking spots were configured within the test area, and their coordinates and occupancy statuses were manually stored in a central parking service database accessible by connecting to the provided backend. The deployment and update process for vehicle control software was validated using Snapcraft. Multiple revisions of the AVP application were published to the Snap Store under the latest/edge channel. The deployment was monitored through version tracking and update history.
The deployed Snap package was installed on the TurtleBot HPC and a cloud backend. For the latter, the correct update process was verified by accessing the REST API endpoint localhost:8088/FreeParkingPlaces, confirming successful backend operation after the update. Additionally, automatic OTA updates were observed on the system using standard Snap tooling. The tool confirms that the software was correctly registered, versioned, and automatically updated via Snap's refresh mechanism.

\subsection{Updating AI Models}
To evaluate the AI software pipeline, a camera-based object detection for the TurtleBots was developed. The AVP lab setup was augmented with obstacles such as wooden blocks and colored cones to simulate realistic perception challenges. Three TurtleBots participated in the evaluation, with one designated as the Ego TurtleBot for deploying the trained AI model as a Docker container. Two software versions were tested: Version 1 detected wooden blocks and other robots, while Version 2 extended detection capabilities to include cones.

\begin{figure}[htp]
    \centering
    \begin{subfigure}[b]{0.48\linewidth}
        \centering
        \includegraphics[width=\linewidth]{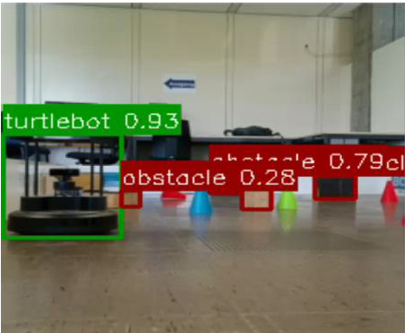}
        \caption{AI Software Version 1}
        \label{fig:ai_version1}
    \end{subfigure}
    \hfill
    \begin{subfigure}[b]{0.48\linewidth}
        \centering
        \includegraphics[width=\linewidth]{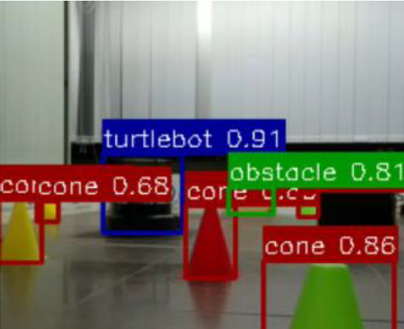}
        \caption{AI Software Version 2}
        \label{fig:ai_version2}
    \end{subfigure}
    \caption{Object detection results from two software versions}
    \label{fig:ai_versions_comparison}
\end{figure}

The object detection capability of the AI software was validated using a YOLOv8 model trained to detect objects in the TurtleBot’s environment. Fig.-\ref{fig:ai_version1} illustrates the detection performance of Software Version 1, demonstrating accurate object localization and classification on the validation dataset. The trained model was containerized as a ROS2 node and deployed on the TurtleBot HPC using Docker.

Deployment was managed via DockerHub, with versioned containers published to a private repository. The middleware, OTA Update Server confirmed update availability through the Flask endpoint. When the endpoint was queried, the server returned the relevant update metadata in JSON format, verifying that the AI software deployment followed the entire CI/CD pipeline. 
The working of the updated software is shown in Fig.~\ref{fig:ai_version2} where the object detection system now identifies the TurtleBot, cones, and obstacles.

\section{Conclusion} \label{sec:conclusion}
An open-source CI/CD pipeline is essential for stable, variant-aware deployment of connected vehicle software across cloud, edge, and in-vehicle systems. The architecture for SDVs is modular and scalable, enabling variant-specific pipelines that ensure compatibility with diverse vehicle configurations. Software is built, tested, versioned, and stored in dedicated repositories, while OTA middleware efficiently manages updates and rollbacks across ECUs, HPCs, and backend systems.

Evaluations show that the OTA middleware reliably handles variant-specific updates and rollbacks, with Snapcraft containerization ensuring stable versioning and automatic updates for vehicle control software. Additionally, containerized YOLOv8 object detection on TurtleBots and experiments with autonomous parking and obstacle detection using UGVs confirm the scalable and reliable integration of this approach for future applications.
The key contributions can be summarized as follows:

\begin{itemize}
    \item Development of a modular, open-source CI/CD pipeline that enables stable and variant-specific software deployment for SDVs.
    \item Integration of diverse software types (firmware, vehicle control software, AI models) using dynamic OTA updates and rollback mechanisms.
    \item Validation in a realistic test environment that demonstrates the scalability and practical viability of the approach for heterogeneous hardware platforms.
\end{itemize}

In future work, this pipeline will be deployed to a bigger testbed for connected vehicles, which is currently under construction. Additionally, the capabilities for variant deployment will be further extended, enabling a seamless integration of feature models and variant management tools.


\bibliographystyle{IEEEtran}
\bibliography{bib}

\begin{thebibliography}{10}
\providecommand{\url}[1]{#1}
\csname url@samestyle\endcsname
\providecommand{\newblock}{\relax}
\providecommand{\bibinfo}[2]{#2}
\providecommand{\BIBentrySTDinterwordspacing}{\spaceskip=0pt\relax}
\providecommand{\BIBentryALTinterwordstretchfactor}{4}
\providecommand{\BIBentryALTinterwordspacing}{\spaceskip=\fontdimen2\font plus
\BIBentryALTinterwordstretchfactor\fontdimen3\font minus \fontdimen4\font\relax}
\providecommand{\BIBforeignlanguage}[2]{{%
\expandafter\ifx\csname l@#1\endcsname\relax
\typeout{** WARNING: IEEEtran.bst: No hyphenation pattern has been}%
\typeout{** loaded for the language `#1'. Using the pattern for}%
\typeout{** the default language instead.}%
\else
\language=\csname l@#1\endcsname
\fi
#2}}
\providecommand{\BIBdecl}{\relax}
\BIBdecl

\bibitem{goswami2024sdv}
\BIBentryALTinterwordspacing
P.~Goswami, ``The software-defined vehicle and its engineering evolution: Balancing issues and challenges in a new paradigm of product development,'' SAE International, Tech. Rep. EPR2024007, 2024, accessed: 2025-05-06. [Online]. Available: \url{https://doi.org/10.4271/EPR2024007}
\BIBentrySTDinterwordspacing

\bibitem{dettinger2024function}
F.~Dettinger, M.~Weiß, and M.~Weyrich, ``Future use cases for vehicular communication based on connected functions,'' in \emph{2024 IEEE 100th Vehicular Technology Conference (VTC2024-Fall)}, 2024, pp. 1--5.

\bibitem{madhuri2025sdv}
T.~S. Madhuri and J.~Bala~Vishnu, ``Software-defined vehicles: The future of automobile industry,'' in \emph{2025 17th International Conference on COMmunication Systems and NETworks (COMSNETS)}, 2025, pp. 192--197.

\bibitem{dettinger2024survey}
F.~Dettinger, M.~Wei{\ss}, D.~Dittler, J.~St{\"u}mpfle, M.~Artelt, and M.~Weyrich, ``A survey on performance, current and future usage of vehicle-to-everything communication standards,'' \emph{arXiv preprint arXiv:2410.10264}, 2024.

\bibitem{weiss2023continuous}
M.~Wei{\ss}, M.~M{\"u}ller, F.~Dettinger, N.~Jazdi, and M.~Weyrich, ``Continuous analysis and optimization of vehicle software updates using the intelligent digital twin,'' in \emph{2023 IEEE 28th International Conference on Emerging Technologies and Factory Automation (ETFA)}.\hskip 1em plus 0.5em minus 0.4em\relax IEEE, 2023, pp. 1--7.

\bibitem{seidel2024variant}
L.~Seidel, H.~Guissouma, A.~Schmid, and E.~Sax, ``\BIBforeignlanguage{english}{Variant-aware reconfiguration of automotive virtual test environments},'' in \emph{\BIBforeignlanguage{english}{Vol.9 - Driving Simulation Conference Europe 2024 VR (DSC 2024), Driving Simulation and Virtual Reality Conference and Exhibition, Straßburg, 18th-20th September 2024}}, 2024, pp. 221 -- 226.

\bibitem{weiss2024simulating}
\BIBentryALTinterwordspacing
M.~Weiß, J.~St\"{u}mpfle, F.~Dettinger, N.~Jazdi, and M.~Weyrich, ``Simulating cloud environments of connected vehicles for anomaly detection,'' in \emph{SAE Technical Paper Series}, ser. STUT.\hskip 1em plus 0.5em minus 0.4em\relax SAE International, Jul. 2024. [Online]. Available: \url{http://dx.doi.org/10.4271/2024-01-2996}
\BIBentrySTDinterwordspacing

\bibitem{federate2024roadmap}
\BIBentryALTinterwordspacing
{FEDERATE Consortium}, ``European software-defined vehicle of the future (sdvof) initiative – vision and roadmap,'' Apr. 2024, accessed: 2025-05-06. [Online]. Available: \url{https://federate-sdv.eu/2024/04/12/european-software-defined-vehicle-of-the-future-sdvof-initiative-vision-and-roadmap/}
\BIBentrySTDinterwordspacing

\bibitem{stoetzner2022deployment}
\BIBentryALTinterwordspacing
M.~Stötzner, S.~Becker, U.~Breitenbücher, K.~Képes, and F.~Leymann, ``Modeling different deployment variants of a composite application in a single declarative deployment model,'' \emph{Algorithms}, vol.~15, no.~10, 2022. [Online]. Available: \url{https://www.mdpi.com/1999-4893/15/10/382}
\BIBentrySTDinterwordspacing

\bibitem{wozniak2015automotive}
L.~Wozniak and P.~Clements, ``How automotive engineering is taking product line engineering to the extreme,'' in \emph{Proceedings of the 19th international conference on software product line}, 2015, pp. 327--336.

\bibitem{stumpfle_tackling_2024}
\BIBentryALTinterwordspacing
J.~Stümpfle, N.~Jazdi, and M.~Weyrich, ``\BIBforeignlanguage{en}{Tackling {Erosion} in {Variant}-{Rich} {Software} {Systems}: {Challenges} and {Approaches}},'' \emph{\BIBforeignlanguage{en}{Procedia CIRP}}, vol. 128, pp. 633--637, 2024. [Online]. Available: \url{https://linkinghub.elsevier.com/retrieve/pii/S221282712400742X}
\BIBentrySTDinterwordspacing

\bibitem{knieke_managed_2022}
\BIBentryALTinterwordspacing
C.~Knieke, A.~Rausch, M.~Schindler, A.~Strasser, and M.~Vogel, ``\BIBforeignlanguage{en}{Managed {Evolution} of {Automotive} {Software} {Product} {Line} {Architectures}: {A} {Systematic} {Literature} {Study}},'' \emph{\BIBforeignlanguage{en}{Electronics}}, vol.~11, no.~12, p. 1860, Jan. 2022, number: 12 Publisher: Multidisciplinary Digital Publishing Institute. [Online]. Available: \url{https://www.mdpi.com/2079-9292/11/12/1860}
\BIBentrySTDinterwordspacing

\bibitem{10710832}
J.~Stümpfle, S.~Baum, D.~Dittler, N.~Jazdi, and M.~Weyrich, ``Automating software product line adoption based on feature models using large language models,'' in \emph{2024 IEEE 29th International Conference on Emerging Technologies and Factory Automation (ETFA)}, 2024, pp. 1--4.

\bibitem{11024511}
J.~Stümpfle, D.~Atray, N.~Jazdi, and M.~Weyrich, ``Large language model assisted transformation of software variants into a software product line,'' in \emph{2025 IEEE/ACM 22nd International Conference on Software and Systems Reuse (ICSR)}, 2025, pp. 12--20.

\bibitem{noauthor_autosar_nodate}
\BIBentryALTinterwordspacing
{AUTOSAR Consortium}, ``{AUTOSAR} {Adaptive} {Platform}.'' [Online]. Available: \url{https://www.autosar.org/standards/adaptive-platform}
\BIBentrySTDinterwordspacing

\bibitem{weiss2022devops}
M.~Weiß, F.~Dettinger, N.~Jazdi, and M.~Weyrich, ``Devops als enabler der kontinuierlichen funktionsverbesserung und automatisierten update-analyse in software-definierten systemen,'' in \emph{Automation 2023}, 2023.

\bibitem{ferreira2025mlops}
A.~L. Ferreira and J.~M. Fernandes, ``Mlops for developing machine-learning-enhanced automotive applications: Experience at the bosch cross-domain computing solutions division,'' \emph{IEEE Software}, vol.~42, no.~1, pp. 34--41, 2025.

\bibitem{covesa2025web}
\BIBentryALTinterwordspacing
{COVESA}, ``Connected vehicle systems alliance (covesa),'' 2025, accessed: 2025-05-06. [Online]. Available: \url{https://covesa.global/}
\BIBentrySTDinterwordspacing

\bibitem{soafee2025doc}
\BIBentryALTinterwordspacing
{SOAFEE Project}, ``Soafee architecture documentation,'' 2025, accessed: 2025-05-06. [Online]. Available: \url{https://architecture.docs.soafee.io/en/latest/contents/architecture.html}
\BIBentrySTDinterwordspacing

\bibitem{weiss2024ad}
M.~Weiß, S.~Thich, M.~Artelt, and M.~Weyrich, ``A survey about self-adaptive anomaly-detection in software-defined systems,'' in \emph{IECON 2024 - 50th Annual Conference of the IEEE Industrial Electronics Society}, 2024, pp. 1--4.

\end{thebibliography}

\end{document}